\documentclass[10pt,twocolumn,letterpaper]{article}

\usepackage{wacv}              

\usepackage{graphicx}
\usepackage{amsmath}
\usepackage{amssymb}
\usepackage{booktabs}
\usepackage{tabularx}
\usepackage{multirow}
\usepackage{bm}

\usepackage[pagebackref,breaklinks,colorlinks]{hyperref}

\usepackage[capitalize]{cleveref}
\crefname{section}{Sec.}{Secs.}
\Crefname{section}{Section}{Sections}
\Crefname{table}{Table}{Tables}
\crefname{table}{Tab.}{Tabs.}

\newcommand{\beginsupplement}{%
        \setcounter{table}{0}
        \renewcommand{\thetable}{S\arabic{table}}%
        \setcounter{figure}{0}
        \renewcommand{\thefigure}{S\arabic{figure}}%
     }
\makeatletter
\newcommand{\printfnsymbol}[1]{%
	\textsuperscript{\@fnsymbol{#1}}%
}
\makeatother
\def\wacvPaperID{1738} 
\def\confName{WACV}
\def\confYear{2025}

\begin{document}

\title{MAISI: Medical AI for Synthetic Imaging}


\author{
{Pengfei Guo$^{1}$\thanks{Equal contribution. The code is available at \href{https://github.com/NVIDIA-Medtech/NV-Generate-CTMR}{NVIDIA MedTech · Open Models Hub}. The online demo is available at \href{https://build.nvidia.com/nvidia/maisi}{NVIDIA NIM}.}   \qquad  Can Zhao$^{1*}$  \qquad Dong Yang$^{1*}$ \qquad  Ziyue Xu$^{1}$} \\
{Vishwesh Nath$^{1}$ \qquad  Yucheng Tang$^{1}$ \qquad Benjamin Simon$^{2}$ \qquad Mason Belue$^{3}$} \\
{Stephanie Harmon$^{2}$ \qquad Baris Turkbey$^{2}$ \qquad Daguang Xu$^{1}$}\\
\vspace{0.4em} 
{$^{1}$NVIDIA \quad $^{2}$National Institutes of Health \quad $^{3}$University of Arkansas for Medical Sciences}\\
\vspace{0.4em} 
}


\maketitle

\begin{abstract}
   Medical imaging analysis faces challenges such as data scarcity, high annotation costs, and privacy concerns. This paper introduces the Medical AI for Synthetic Imaging (MAISI), an innovative approach using the diffusion model to generate synthetic 3D computed tomography (CT) images to address those challenges. MAISI leverages the foundation volume compression network and the latent diffusion model to produce high-resolution CT images (up to a landmark volume dimension of 512 $\times$ 512 $\times$ 768 ) with flexible volume dimensions and voxel spacing. By incorporating ControlNet, MAISI can process organ segmentation, including 127 anatomical structures, as additional conditions and enables the generation of accurately annotated synthetic images that can be used for various downstream tasks. Our experiment results show that MAISI's capabilities in generating realistic, anatomically accurate images for diverse regions and conditions reveal its promising potential to mitigate challenges using synthetic data.
\end{abstract}

\section{Introduction}
\label{sec:intro}

\begin{figure}[t] 
\centering
\includegraphics[width=\columnwidth]{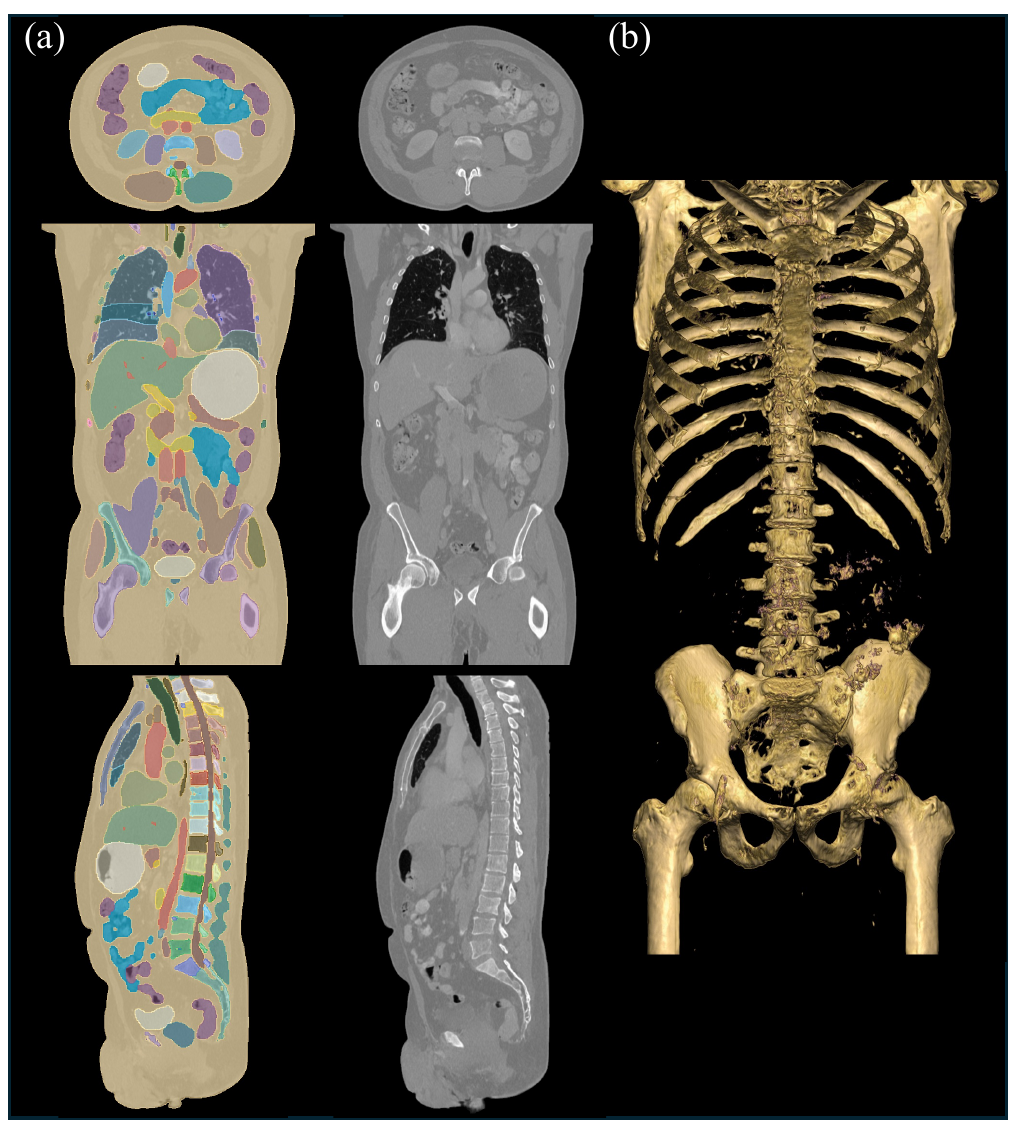}
\caption{(a) A generated high-resolution CT volume (with volume dimensions of 512 $\times$ 512 $\times$ 768 and voxel spacing of 0.86 $\times$ 0.86 $\times$ 0.92 $\text{mm}^3$) by the proposed method and its corresponding segmentation condition overlaid on generated volume. We show the axial, sagittal, and coronal views from top to bottom, respectively. (b) 3D volume rendering of generated CT by MAISI. The rendering setting is tuned to highlight bone structures and demonstrate the realism of the generated CT volume.}
\label{fig:highres_result}
\end{figure}

Medical imaging analysis has been integral to modern healthcare, providing critical insights into patient diagnosis, treatment planning, and monitoring. The rapid advancement of machine learning (ML) approaches has revolutionized diagnostic and therapeutic practices in modern healthcare. However, the development of effective ML models in this domain continues to face the following significant challenges~\cite{kang2023label,liu2023clip,zhou2019high}: (1) \textbf{data scarcity}: the rarity of certain medical conditions (\eg, certain types of cancer, and rare diseases) complicates the data acquisition process, which leads to the limited acquired data that might not adequately represent the diversity of real-world cases. (2) \textbf{high human-annotation costs}: annotating medical images, such as MRI and CT scans, is inherently more expertise-demanding than annotating objects in general images. Medical images often contain subtle features that are critical for accurate diagnosis and treatment. Expert knowledge is usually required to accurately identify and annotate these conditions.
(3) \textbf{privacy concerns}: conventional data acquisition and processing of medical images often require access to large volumes of patient data, which raises ethical concerns and poses significant logistical challenges due to the sensitive nature of patient information.

To address these limitations, generating synthetic data has emerged as a promising direction. By creating artificial yet realistic medical images, synthetic data can augment existing datasets, reduce the dependency on real patient data, and provide a cost-effective alternative to manual data annotation. With the recent advancement of the generative model, many novel approaches, such as generative adversarial networks (GAN)~\cite{goodfellow2014generative} and Diffusion Models (DM)~\cite{ho2020denoising}, have been extensively studied for their capacity to generate photo-realistic images in various tasks in general computer vision society. In the context of medical image generation, several generative models have been successfully
applied for medical image synthesis, such as multi-contrast MR/CT image synthesis~\cite{joyce2017robust,guo2020anatomic,sun2022hierarchical}, cross-modality image translation~\cite{chartsias2017multimodal, shin2018medical,zhao2017whole,yang2020unsupervised,dewey2019deepharmony}, and image reconstruction~\cite{peng2022towards, xie2022measurement,zhao2020smore,darestani2024ir}. 

However, several key challenges are not fully explored in previous studies. \textbf{First}, realistic high-resolution (larger than the volume dimension of 512$^3$) 3D volume generation is still a challenging task due to the huge memory consumption imposed by unified 3D frameworks, which must handle the vast amount of data involved in such high-dimensional representations~\cite{singh20203d}. Overcoming this memory bottleneck is essential for advancing the realism and applicability of 3D volume generation in clinical contexts. \textbf{Second}, the constraint of fixed output volume dimensions and voxel spacing poses substantial limitations in real-world applications~\cite{chen2021deep}. These parameter presets are often incompatible with the diverse requirements of different tasks, such as the analysis of varying anatomical structures. The ability to dynamically adjust both the volume dimensions and the voxel spacing is crucial for enhancing the flexibility and utility of 3D generative models.
\textbf{Third}, another common limitation of current generative models for medical image generation is their specialization to dedicated datasets or particular types of organs. These models, once trained, are typically not generalizable beyond the specific data and target organ they are developed on, which restricts their broader application in diverse settings. Developing more versatile models that can adapt to multiple datasets and organ types and mitigate the need for extensive retraining is a key objective for advancing the field~\cite{chen2024towards}.

In this paper, we propose a method, namely Medical AI for Synthetic Imaging (MAISI), a new framework for high-resolution 3D CT volume generation, which consists of three 3D networks including two foundation models (\ie, a volume compression network, a latent diffusion model~\cite{rombach2022high}) and a ControlNet~\cite{zhang2023adding} for versatile generation tasks. Volume Compression Network is trained on a large amount of data (\ie, 39,206 3D CT volumes) and is responsible for compressing the 3D medical images into latent space and mapping the generated latent features back to image space by a visual encoder and a visual decoder, respectively. To reduce the memory footprint, we introduce the tensor splitting parallelism (TSP) inspiring from the tensor parallelism technique~\cite{shoeybi2019megatron}, originally proposed for linear layers, to the 3D convolutional layers allowing for the encoding and decoding of high-resolution CT volumes in a unified 3D network.
The latent diffusion model in MAISI facilitates the creation of realistic latent features of 3D medical images. Benefiting from a compressed latent space with flexible dimensions and taking body region and voxel spacing as conditions, it enables the generation of complex anatomical structures with a high degree of fidelity while maintaining relatively low memory consumption. The latent diffusion model is trained on 10,277 CT volumes from diverse datasets, encompassing various body regions and disease conditions to enhance its generalizability and robustness, which enables the model to capture the knowledge represented in a wide range of clinical scenarios. Further, the integration of ControlNet~\cite{zhang2023adding} into the MAISI framework introduces a mechanism for dynamic control over the generated outputs. This component enhances MAISI's versatility and applicability across a wider range of tasks (\eg, conditional generation based on segmentation masks, as illustrated in Fig.~\ref{fig:highres_result}, image inpainting, \etc.). Additionally, this capability minimizes the need for extensive retraining of the two underlying foundation models when transitioning between different tasks or clinical objectives, thereby conserving both time and computational resources.

To summarize, this paper makes the following contributions:
\begin{itemize}
  \item A novel framework, MAISI, for high-resolution 3D CT volume generation is proposed, which enables the versatile generation of synthetic CT images.
  \item Tensor splitting parallelism (TSP) is introduced to 3D convolutional networks. To the best of our knowledge, MAISI is the first attempt to generate realistic 3D CT images larger than 512$^3$ voxels. 
  \item MAISI provides dynamic control over outputs, enabling annotated synthetic images to improve downstream task performance.
\end{itemize}
\begin{figure*}[t] 
\centering
\includegraphics[width=\textwidth]{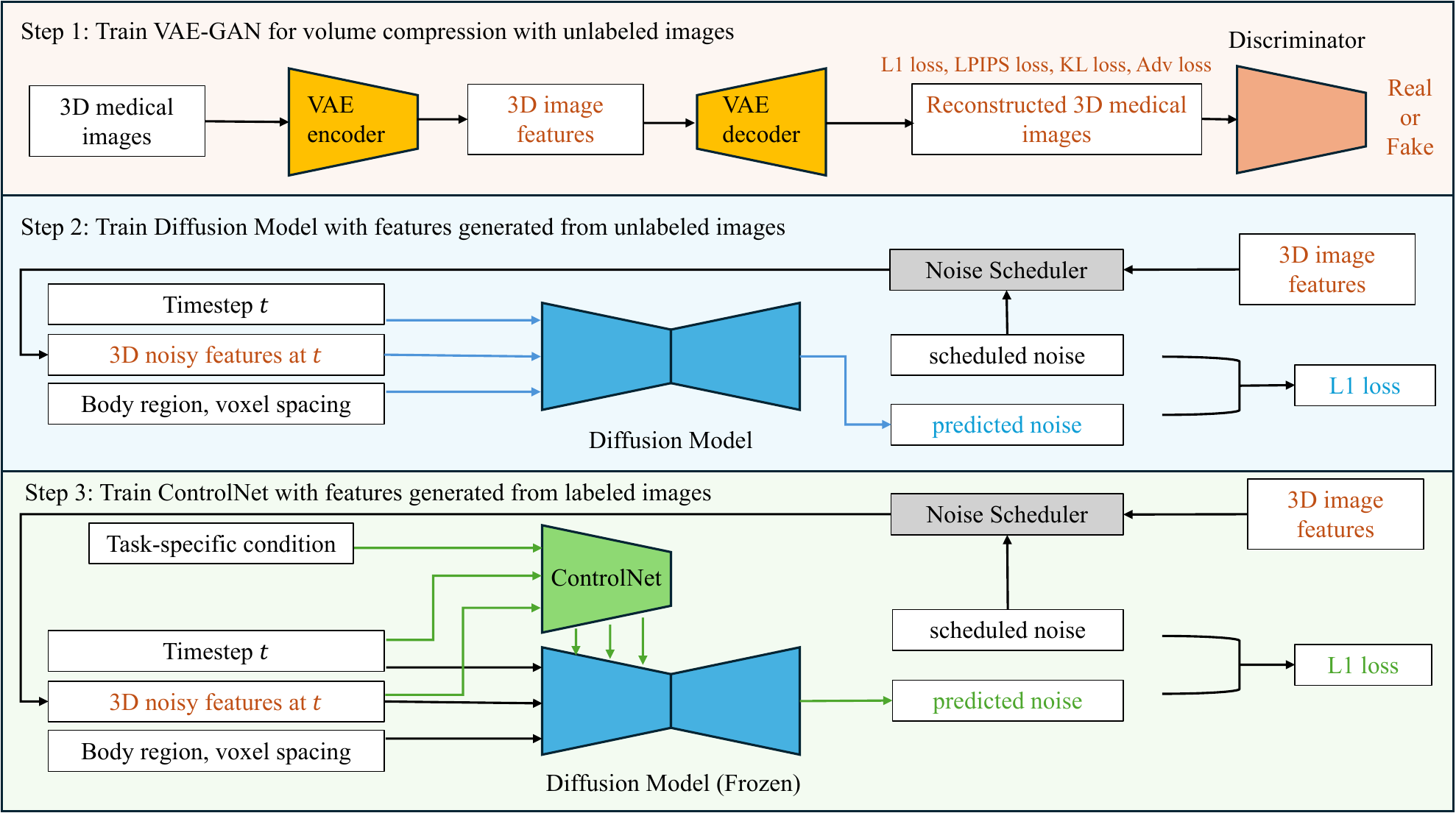}
\caption{The overview of three development stages of MAISI.}
\label{fig:training_stage}
\end{figure*}

\section{Related Work}
Medical image synthesis has become an increasingly prominent research area, particularly in response to the challenges discussed in Sec.~\ref{sec:intro}. Early approaches~\cite{roy2011compressed,miller1993mathematical,cardoso2015template} to medical image synthesis were predominantly based on traditional image processing techniques, such as the example-based approach~\cite{roy2013magnetic} and geometry-regularized dictionary learning~\cite{huang2016geometry}, which, while effective to some extent, are limited in their ability to generate realistic and diverse medical images. The advent of machine learning, particularly deep learning, has significantly advanced the field, enabling more sophisticated and accurate models for image synthesis~\cite{skandarani2023gans}.

\noindent\textbf{GAN in medical image synthesis.} GAN~\cite{goodfellow2014generative}, one of generative models~\cite{du2019implicit,kingma2019introduction,creswell2018generative,papamakarios2021normalizing}, has been widely adopted for various tasks, such as MRI/CT image synthesis~\cite{joyce2017robust,guo2020anatomic,sun2022hierarchical}, cross-modality image translation~\cite{chartsias2017multimodal, shin2018medical, yurt2021mustgan,yang2020unsupervised}, image reconstruction~\cite{peng2022towards, xie2022measurement} and super-resolution~\cite{you2019ct,pham2019multiscale,peng2020saint,ahmad2022new}, in medical imaging synthesis due to its promising ability to generate high-quality images. One of the most critical applications of GAN in medical imaging is data augmentation by generating annotated images. Several studies~\cite{chartsias2017adversarial,guo2020anatomic,zhang2018translating,huo2018synseg}, have employed GAN to generate lesion images to augment training data for improving downstream tasks to overcome data scarcity issues.
However, those methods focus on 2D medical imaging or small volumetric patch synthesis, which is fundamentally limited due to neglecting the inherent complexity and the 3D nature of medical data. In this work, we focus on generating full CT volume in realistic dimensions (up to 512 $\times$ 512 $\times$ 768) to model complex volumetric features in a unified framework.

\noindent\textbf{DM in medical image synthesis.} Diffusion models~\cite{ho2020denoising, rombach2022high} have recently emerged as a powerful generative model that has shown great potential in medical imaging synthesis due to its capabilities in high-quality image synthesis, stable training process, and flexibility in conditioning~\cite{croitoru2023diffusion,yang2023diffusion,cao2024survey}. ~\cite{konz2024anatomically,shang2023synfundus,khosravi2024synthetically} demonstrate the effectiveness of DM-based methods in generating high-quality 2D medical images that capture intricate details with minimal artifacts, making them suitable for clinical use. GenerateCT~\cite{hamamci2023generatect} is designed to synthesize 3D CT volumes from free-form medical text prompts and accomplishes arbitrary-size CT volume generation by decomposing the process into a sequential generation of individual slices using DM. However, due to the nature of 2D approaches, the issue of 3D structural inconsistencies across slices is noticeable and problematic in the generated images. Application-wise, many recent studies~\cite{lai2024pixel,wu2024freetumor,hu2022synthetic,lyu2022learning,zhang2023unsupervised,zhang2023self} are focusing on tumor synthesizing and improving models' performance in downstream tasks. DiffTumor~\cite{chen2024towards} seeks to enhance the robustness and generalizability of tumor segmentation models across various organs, such as the liver, pancreas, and kidney, by leveraging high-quality synthetic tumors generated through specialized diffusion models. In this work, we focus on achieving conditional generation tailored to versatile tasks by leveraging robust foundation models, which significantly minimizes the need for extensive retraining across different applications, thereby conserving both time and computational resources while maintaining adaptability and efficiency in diverse clinical scenarios.

\section{Methodology}
As shown in Fig.~\ref{fig:training_stage}, the development of MAISI
involves three stages. In the first stage, the volume compression network (\ie, VAE-GAN~\cite{rombach2022high}) is trained on a substantial dataset comprising 39,206 3D CT volumes and 18,827 3D MRI volumes. This network effectively compresses high-resolution 3D medical images into a latent space that is perceptually equivalent to the image space, reducing memory usage and computational complexity for later stages. In the second stage, a latent diffusion model is trained on 10,277 CT volumes sourced from diverse datasets. This model operates within the compressed latent space, conditioned on specific body regions and voxel spacing, to generate features of realistic and complex 3D anatomical structures in flexible dimensions. Training on a broad range of data enhances the model’s generalizability and adaptability in different tasks. The final stage involves the integration of ControlNet~\cite{zhang2023adding} into the MAISI framework. This component allows for dynamic control over the generated outputs by injecting additional conditions into the trained latent DM in the second stage, potentially supporting a wide range of tasks. The integration reduces the need for extensive retraining when the model is adapted to different tasks. In what follows, we provide detailed descriptions of each key component of the MAISI framework.

\subsection{Volume Compression Network}
The volume compression model builds upon previous studies~\cite{rombach2022high,esser2021taming} and employs a Variational Autoencoder (VAE) trained on combined objectives, which integrates perceptual loss $\mathcal{L}_{\text{lpips}}$~\cite{zhang2018unreasonable}, adversarial loss~\cite{yu2021vector} $\mathcal{L}_{\text{adv}}$, and L1 reconstruction loss $\mathcal{L}_{\text{recon}}$ on voxel-space. These combined objectives ensure that the volume reconstructions adhere closely to the image manifold and enforce local realism~\cite{rombach2022high}. In addition, we follow~\cite{kingma2013auto,rombach2022high,rezende2014stochastic} adding Kullback-Leibler (KL) regularization $\mathcal{L}_{\text{reg}}$ toward a standard normal on the learned latent features for avoiding high-variance latent spaces.

Given a CT volume $x \in \mathbb{R}^{H\times W \times D}$ in grayscale voxel
space, where $H$ denotes the height, $W$ the width, and $D$ the
depth, the encoder $\mathcal{E}$ of AE downsamples $x$ and generates the 
latent representation $z=\mathcal{E}(x) \in \mathbb{R}^{h\times w \times d}$ with much smaller spatial dimensions. The decoder $\mathcal{D}$ of AE approximates the reconstructed volume $\tilde{x} = \mathcal{D}(z) = \mathcal{D}(\mathcal{E}(x))$ from the latent features. A 3D discriminator, denoted as 
$\mathcal{C}$, is utilized to identify and penalize any unrealistic artifacts in the reconstructed volume $\tilde{x}$. As shown in Fig.~\ref{fig:training_stage} step 1, the overall objective $\mathcal{L}_{\text{AE}}$ to train the volume compression network ($\mathcal{E}, \mathcal{D}$) in MAISI can be defined as follows:
\setlength{\belowdisplayskip}{2pt} \setlength{\belowdisplayshortskip}{2pt}
\setlength{\abovedisplayskip}{2pt} \setlength{\abovedisplayshortskip}{2pt}
\begin{equation}
\begin{aligned}
     \min\limits_{\mathcal{E}, \mathcal{D}}\max\limits_{\mathcal{C}}\Bigl(\mathcal{L}_{\text{recon}}(x,\mathcal{D}(\mathcal{E}(x))) &+ 
    \mathcal{L}_{\text{lpips}}(x,\mathcal{D}(\mathcal{E}(x))) \\
     &+ \mathcal{L}_{\text{reg}}(\mathcal{E}(x)) + \mathcal{L}_{\text{adv}} \Bigr),
\end{aligned}
\end{equation}
where 
\begin{equation}
\begin{aligned}
    \mathcal{L}_{\text{adv}} = \log\mathcal{C}(x) + \log(1-\mathcal{C}(\mathcal{D}(\mathcal{E}(x)))).
\end{aligned}
\end{equation}

\begin{figure}[t] 
\centering
\includegraphics[width=\columnwidth]{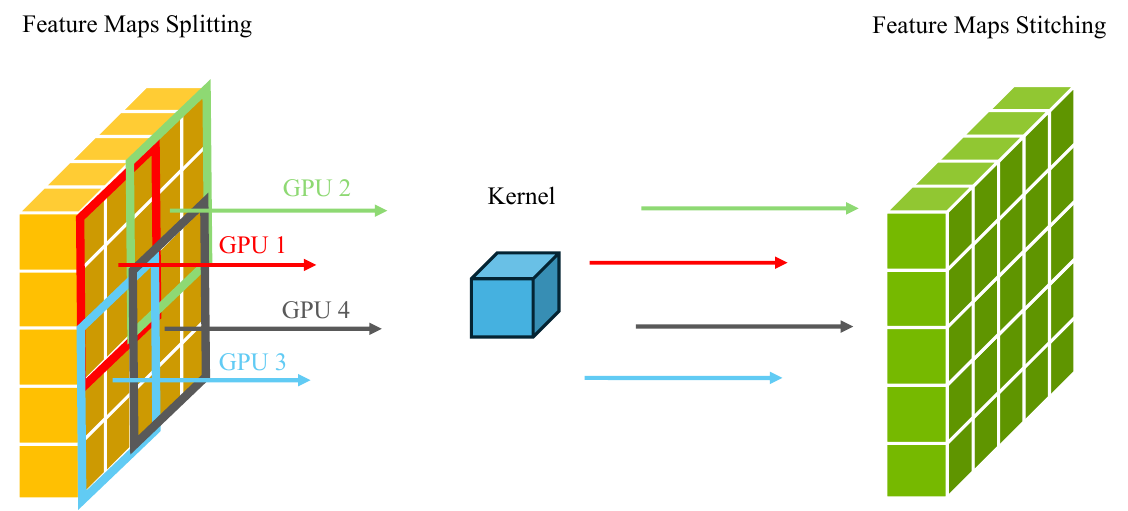}
\caption{The schematics of tensor splitting parallelism in MAISI. Feature maps are first partitioned into smaller segments with overlaps and allocated to designated devices. Then, these segments are stitched together to compose the output of the layer.}
\label{fig:tp}
\end{figure}

Generating high-resolution 3D volumes, particularly those exceeding dimensions of $512^3$ voxels, poses a significant challenge due to the substantial memory demands imposed by the 3D convolution networks. In order to address memory bottleneck, previous methods~\cite{hamamci2023generatect,saharia2022photorealistic} achieve 2D high-resolution image synthesis via an additional super-resolution model. However, in the context of 3D whole-volume generation, the memory consumption can still quickly reach the hardware limitation of modern GPUs (\eg, NVIDIA A100 80G). To overcome GPU memory constraints, sliding window inference~\cite{cardoso2022monai} is a common technique. It divides the large network input into smaller 3D patches in a sliding-window fashion and then stitches the network's output of each patch together to form the final results. When used in the 3D medical image segmentation model inference, it can often lead to artifacts/discontinuities along the window boundaries. While overlapping windows can help in segmentation tasks by smoothing over the boundary artifacts of probability maps, we empirically found this issue in transition areas between windows is more pronounced for the synthesis task due to the direct generation of image intensities, and thus the direct adaptation of sliding window inference is not self-sufficient. To minimize the use of the sliding-window approach for image synthesis, we propose a simple yet effective solution by introducing tensor splitting parallelism into convolutional networks. The tensor parallelism~\cite{shoeybi2019megatron} is initially developed to distribute the inputs or model weights of matrix multiplication operations in fully connected layers across multiple GPUs. Unlike language models~\cite{dubey2024llama} built upon linear layers, the memory bottleneck usually attributes to the large feature maps. As demonstrated in Fig.~\ref{fig:tp}, the proposed TSP is utilized to divide feature maps into smaller segments while preserving necessary overlaps across both convolution and normalization layers of AE. Each segment is assigned to a designated device, and these segments are subsequently merged to generate the layer's output. This flexible implementation enables segments to be distributed across multiple devices to accelerate the inference and also allows each segment to be processed sequentially within a single device in a loop to reduce peak memory usage.

\subsection{Diffusion Model}\label{sec:dm}

The Diffusion Model in MAISI operates on a compressed latent space with flexible dimensions and incorporates body region and voxel spacing as conditional inputs, facilitating the high-fidelity generation of anatomical structures. Diffusion models are probabilistic models that aim to learn a data distribution $p(x)$ by gradually denoising a normally distributed variable. This process is equivalent to learning the reverse dynamics of a fixed Markov Chain over a sequence of $T$ steps. The denoising score-matching~\cite{song2020score} is widely adopted in image synthesis tasks~\cite{dhariwal2021diffusion,saharia2022image}. In the context of latent diffusion model~\cite{rombach2022high}, the learning model $\epsilon_\theta$ functions as a uniformly weighted sequence of denoising autoencoders $\epsilon_\theta(z_t, t); t= 1\dots T$, which are designed to predict a denoised version of the input latent features $z_t$ and $z_t$ represents a noisy variant of the original input at time step $t$. The neural backbone $\epsilon_\theta$ is defined as a time-conditional U-Net~\cite{ronneberger2015u}

As shown in Fig.~\ref{fig:training_stage} step 2, the diffusion model in MAISI additionally conditions on both the body region and voxel spacing. The body region is defined by a top-region index $\bm{i}_{\text{top}}$ and a bottom-region index $\bm{i}_{\text{bottom}}$, indicating the extent of the CT scan coverage.  $\bm{i}_{\text{top}}$ and $\bm{i}_{\text{bottom}}$ are defined by 4-dimensional one-hot vectors for head-neck, chest, abdomen, and lower-body regions). We ascertain the body region either through segmentation ground truth or predated segmentation masks from whole-body CT segmentation models, such as TotalSegmentator~\cite{wasserthal2023totalsegmentator} or VISTA3D~\cite{he2024vista3d}. The condition of voxel spacing $\bm{s}$ is defined by a vector containing three float numbers representing the physical size of each voxel along each of the three dimensions in millimeters. We denote the primary conditions as $\bm{c}_p:=\{\bm{i}_{\text{top}}, \bm{i}_{\text{bottom}}, \bm{s}\}$.  Formally, the training objective of MAISI diffusion model is as follows:
\begin{equation}
\begin{aligned}
    \mathbb{E}_{\mathcal{E}(x),\epsilon\sim\mathcal{N}(0,1),t, \bm{c}_p} \Bigl[\lVert \epsilon - \epsilon_\theta(z_t, t, \bm{c}_p) \rVert_1\Bigr],
\end{aligned}
\end{equation}
where the neural backbone $\epsilon_\theta$ is configured to condition on time step $t$ and the primary conditions as $\bm{c}_p$. Moreover, $\epsilon_\theta$ undergoes training on the latent variable $z_t$, which varies in dimensions throughout the training process. This training regimen is designed to facilitate the generation of outputs with flexible volumetric dimensions.

\subsection{Additional Conditioning Mechanisms}\label{sec: controlnet}

In addition to the primary conditioning on body region and voxel spacing described in the Sec.~\ref{sec:dm}, MAISI incorporates an additional mechanism for enhancing the control and flexibility of the generated outputs through the integration of ControlNet~\cite{zhang2023adding}. It is seamlessly embedded into the MAISI architecture with the latent diffusion model, to provide additional conditioning paths that allow for task-specific adaptations. ControlNet~\cite{zhang2023adding} is designed to inject auxiliary conditions into the diffusion process, enabling more precise control over the generated anatomical structures. It operates by creating two copies of the neural network blocks: a locked copy that preserves the original model's knowledge, and a trainable copy that learns to respond to specific conditions. These copies are connected using zero convolution layers, which gradually evolve from zero weights to optimal settings during training. These additional conditions can include a variety of inputs such as segmentation masks for conditional generation based on masks, or masked images and tumor masks for the tumor inpainting~\cite{chen2024towards}. Similar to~\cite{zhang2023adding, konz2024anatomically, mou2024t2i}, we employ a compact encoder network to transform the additional condition from its original resolution into latent features, which are denoted by the task-specific condition $\bm{c}_f$. This transformation process effectively aligns the additional condition with the spatial dimensions of the latent space.
The integration of ControlNet~\cite{zhang2023adding}  occurs during the third stage (Fig.~\ref{fig:training_stage} step 3) of MAISI's development, where it is trained with the frozen latent diffusion model. The overall learning objective of the entire diffusion algorithm, which incorporates the ControlNet~\cite{zhang2023adding}, is formulated as follows:
\begin{equation}
\begin{aligned}
    \mathbb{E}_{\mathcal{E}(x),\epsilon\sim\mathcal{N}(0,1),t, \bm{c}_p,  \bm{c}_f} \Bigl[\lVert \epsilon - \epsilon_\theta(z_t, t, \bm{c}_p,  \bm{c}_f) \rVert_1\Bigr].
\end{aligned}
\end{equation}
This integration adds a flexible mechanism to MAISI for controlling the generation of 3D anatomical structures. By injecting task-specific conditions, MAISI can be fine-tuned to meet the specific needs of various medical imaging tasks without retraining the two foundation models, making it a versatile tool for various medical image synthesis tasks.
\section{Experiments}

\subsection{Datasets and Implementation Details}\label{sec: data}

To develop and evaluate the proposed MAISI framework, we curate a large-scale medical imaging dataset from publicly available datasets to capture a diverse range of anatomical structures, imaging conditions, and disease states. These datasets are integral to training the three networks within the MAISI framework. 
The \textbf{Volume Compression Network} (MAISI VAE) is trained on a dataset comprising 37,243 CT volumes for training and 1,963 CT volumes for validation, covering the chest, abdomen, and head and neck regions. Additionally, we include 17,887 MRI volumes for training and 940 MRI volumes for validation, spanning the brain, skull-stripped brain, chest, and below-abdomen regions to potentially support MRI modality in future work. 
The \textbf{Latent DM} (MAISI Diffusion Model) was trained using 10,277 CT volumes sourced from multiple public datasets. These datasets are chosen to represent various clinical scenarios, including different body regions and pathological conditions. Including diverse voxel spacings and anatomical regions as conditional inputs during training is essential to ensure the model's ability to generate high-fidelity anatomical structures with flexible dimensions. For compatibility with the shape requirement of U-Net~\cite{ronneberger2015u}, we resample the dimensions of volumes to the multiples of 128 in this stage. Supplementary Fig.~\ref{fig:dataset_info} visualizes the characteristics and spatial complexity of the data involved in training the diffusion model. 
The \textbf{ControlNet} part was further trained using subsets of the datasets used for the diffusion model based on different downstream tasks, with additional annotations such as segmentation masks and tumor labels. For example, segmentation masks with 127 anatomical structures are derived from annotated ground truth or pre-trained models, such as TotalSegmentator~\cite{wasserthal2023totalsegmentator} and VISTA3D~\cite{he2024vista3d}. These additional annotations allow ControlNet to provide fine-grained control over the generation process, enabling tasks such as conditional generation from segmentation masks and tumor inpainting. More details about dataset creation for three development stages can be found in Supplementary Sec.~\ref{supp: data}. We implement all networks using PyTorch~\cite{Ansel_PyTorch_2_Faster_2024} and MONAI~\cite{cardoso2022monai}. The models are trained using the NVIDIA V100 and A100 GPUs. We utilize a quality check function to evaluate the generated images used in downstream tasks, 
which is designed to verify that the median Hounsfield Units (HU) intensity values for major organs in the CT images are within the established normal range from training data. More details about model training are provided in Supplementary Sec.~\ref{supp: training}.

\subsection{Evaluation of MAISI VAE}

\begin{table}[ht!]
\centering
\resizebox{0.99\columnwidth}{!}{
\begin{tabular}{c |c|c|c|c|c} 
\hline
 Dataset & Model & LPIPS $\downarrow$ & SSIM $\uparrow$ & PSNR $\uparrow$ & GPU $\downarrow$ \\
\hline\hline
\multirow{2}{*}{MSD Task07} & MAISI VAE & \textbf{0.038} & \textbf{0.978} & \textbf{37.266} & \textbf{0h} \\
\cline{2-6}
& Dedicated VAE & 0.047 & 0.971 & 34.750 &  619h \\ 
\hline\hline
\multirow{2}{*}{MSD Task08} & MAISI VAE & 0.046 & 0.970 & 36.559 & \textbf{0h} \\
\cline{2-6}
& Dedicated VAE & \textbf{0.041} & \textbf{0.973} & \textbf{37.110} &  669h \\ 
\hline\hline
\multirow{2}{*}{Brats18} & MAISI VAE & \textbf{0.026} & \textbf{0.977} & \textbf{39.003} & \textbf{0h} \\
\cline{2-6}
& Dedicated VAE & 0.030 & 0.975 & 38.971 &  672h \\ 
\hline
\end{tabular}
}
\caption{Performance comparison of the MAISI VAE model on out-of-distribution datasets versus dedicated VAE models. The ``GPU" column shows additional GPU hours for training with one 32G V100 GPU.}
\label{tab:vae_results}

\end{table}
To demonstrate the robustness and generalizability of the MAISI VAE model as a foundational model, we test its performance on several out-of-distribution datasets (\ie, unseen during training), including MSD Pancreas Tumor~\cite{antonelli2022medical} (MSD Task07), MSD Hepatic Vessels~\cite{antonelli2022medical} (MSD Task08), and BraTS18~\cite{bakas2018identifying} (post-contrast T1-weighted MRI). Notably, this application required no additional training, resulting in eliminating any associated training costs of GPU hours. For comparison, we also train the dedicated VAE models separately on each dataset using 80\% of the data, with the same data augmentation techniques and hyper-parameters as those employed for the MAISI VAE training, to establish a benchmark for dedicated VAE models. 

The results from testing on the remaining 20\% of the data, shown in Table \ref{tab:vae_results}, revealed that the MAISI VAE model achieved comparable results without additional GPU resource expenditure. This underscores the model's cost-effectiveness and practicality, suggesting its potential to assist the research community in optimizing resource utilization while maintaining the model's performance.

\subsection{Evaluation of MAISI Diffusion Model}

\noindent\textbf{Synthesis quality.} We assess the synthesis quality of the standalone MAISI DM by conducting comparisons with several established baseline methods, including DDPM~\cite{ho2020denoising}, LDM~\cite{rombach2022high}, and HA-GAN~\cite{sun2022hierarchical}. The first evaluation focuses on comparing the fidelity of images generated by our model against those produced by the HA-GAN~\cite{sun2022hierarchical}, utilizing its publicly available trained weights\footnote{https://github.com/batmanlab/HA-GAN}. Given that HA-GAN~~\cite{sun2022hierarchical} specifically targets CT images of the chest region, we curate a collection of chest CT datasets for this analysis, including MSD Lung Tumor~\cite{antonelli2022medical} (MSD Task06), LIDC-IDRI~\cite{hancock2016lung}, and TCIA COVID-19~\cite{harmon2020artificial}. These datasets provide a diverse range of imaging conditions and pathology, enriching the comparative study. We use the Fréchet Inception Distance (FID)~\cite{heusel2017gans} as the metric for evaluating the similarity between the distributions of generated images and real counterparts from varied sources. Table~\ref{tab:dm_fid_release_ckpt} presents the average FID for both real and synthesized images across the three datasets. Notably, the MAISI DM significantly outperforms HA-GAN~\cite{sun2022hierarchical} in all datasets, demonstrating its capability to generate images with a much closer appearance to real data.

\begin{table}[ht!]
\small
\resizebox{0.99\columnwidth}{!}{
\begin{tabular}{cc|c|c|c}
\hline
\multicolumn{2}{c|}{FID $\downarrow$ (Avg.)}                             & MSD Task 06 & LIDC-IDRI & TCIA
COVID-19    \\ \hline\hline
\multicolumn{1}{c|}{\multirow{3}{*}{Real}}      & MSD Task06 & –           & 3.987     & 1.858  \\ \cline{2-5} 
\multicolumn{1}{c|}{}                           & LIDC-IDRI    & 3.987       & –         & 4.744  \\ \cline{2-5} 
\multicolumn{1}{c|}{}                           & TCIA COVID-19          & 1.858       & 4.744     & –      \\ \hline
\multicolumn{1}{l|}{\multirow{2}{*}{Synthesis}} & HA-GAN~\cite{sun2022hierarchical}       & 98.208      & 116.260   & 98.064 \\ \cline{2-5} 
\multicolumn{1}{l|}{}                           & MAISI DM     & \textbf{4.349}      & \textbf{6.200}    & \textbf{8.346} \\ \hline
\end{tabular}
}
\caption{Fréchet Inception Distance of the
MAISI model and the baseline method using its
released checkpoint with multiple public datasets
as the references.}
\label{tab:dm_fid_release_ckpt}
\end{table}
\begin{table}[ht!]
\small
\resizebox{0.99\columnwidth}{!}{
\begin{tabular}{c|c|c|c|c}
\hline
Method                & FID $\downarrow$ (Axial) & FID $\downarrow$ (Sagittal) & FID $\downarrow$ (Coronal) & FID $\downarrow$ (Avg.) \\ \hline\hline
DDPM~\cite{ho2020denoising}                  & 18.524         & 23.696         & 25.604         & 22.608        \\ \hline
LDM~\cite{rombach2022high}                   & 16.853         & 10.191         & 10.093         & 12.379        \\ \hline
HA-GAN~\cite{sun2022hierarchical}                & 17.432         & 10.266         & 13.572         & 13.757        \\ \hline
MAISI DM & \textbf{3.301}          & \textbf{5.838}          & \textbf{9.109}          & \textbf{6.083}         \\ \hline
\end{tabular}
}
\caption{Fréchet Inception Distance across three views between MAISI DM and retrained baseline methods using the unseen dataset autoPET 2023~\cite{gatidis2022whole} as the reference.}
\label{tab:dm_autopet}
\end{table}

\begin{figure}[t] 
\centering
\includegraphics[width=\columnwidth]{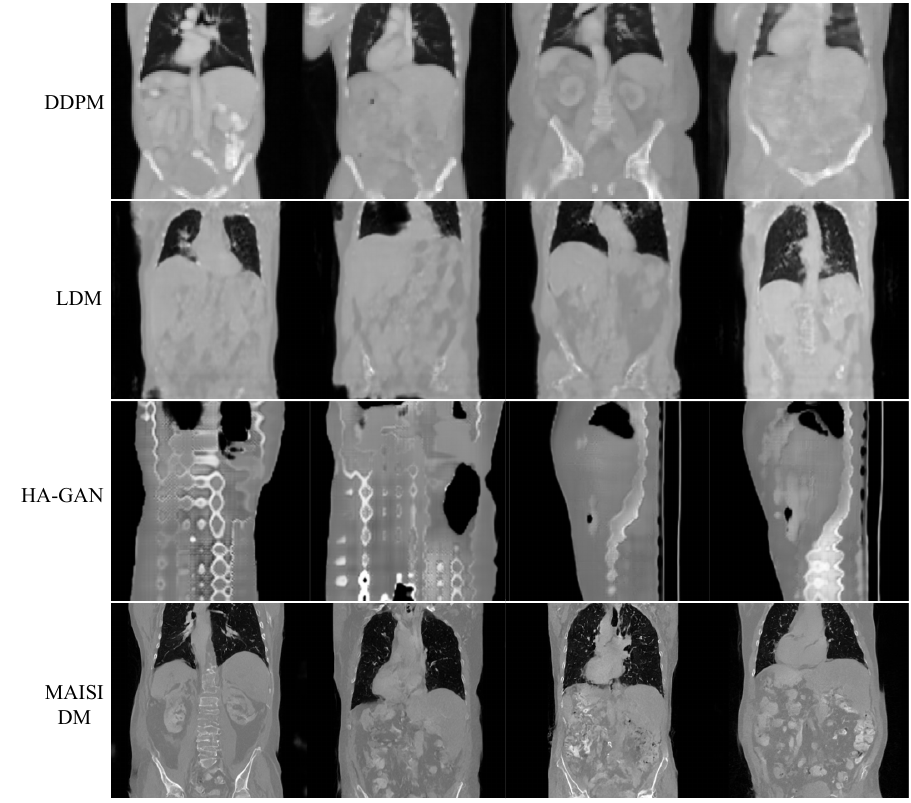}
\caption{Qualitative comparison of generated images between retrained baseline methods using our large-scale datasets and MAISI DM.
}
\label{fig:dm_baselines}
\end{figure}

In addition, we retrain all baseline methods using our large-scale datasets, described in Sec.~\ref{sec: data}. For a more comprehensive evaluation of synthesis quality, we utilize an unseen dataset autoPET 2023~\cite{gatidis2022whole} as the reference to conduct synthesis quality evaluation. This dataset encompasses whole-body CT scans from patients with various types of cancer and negative controls.
Results in Table~\ref{tab:dm_autopet} demonstrate that our MAISI DM surpasses the retrained baseline models in generating high-quality images in the external evaluation. Fig.~\ref{fig:dm_baselines}
presents a visual comparison illustrating that the high-resolution images synthesized by the MAISI DM show improved detail and a more precise representation of global anatomical structures compared to baseline methods. 

\begin{figure}[t] 
\centering

\includegraphics[width=\columnwidth]{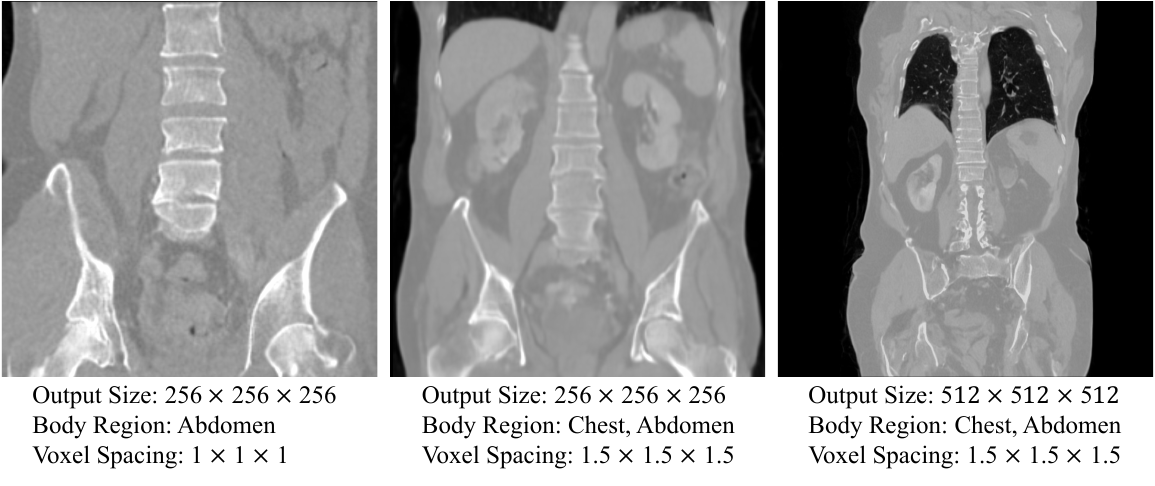}
\caption{The sagittal view of generated CT images from MAISI DM under different primary conditions $\bm{c}_p$. From left to right, the voxel spacing is first increased by 50\%, followed by a doubling of the output dimensions. The coverage of the generated CT images gradually expands, starting from a local region of the abdomen and extending to the entire chest-abdomen region.}
\label{fig:dm_condition_response}
\end{figure}

\noindent\textbf{Response to primary conditions.} Fig.~\ref{fig:dm_condition_response} illustrates the model's adaptability to different body regions and voxel spacing conditions. The MAISI model effectively generates anatomically consistent and high-quality images across different primary conditions $\bm{c}_p$, demonstrating its flexibility and control over synthesized images.

\subsection{Data Augmentation in Downstream Tasks}\label{sec: da}
\begin{figure*}[t] 
\centering
\includegraphics[width=0.95\textwidth]{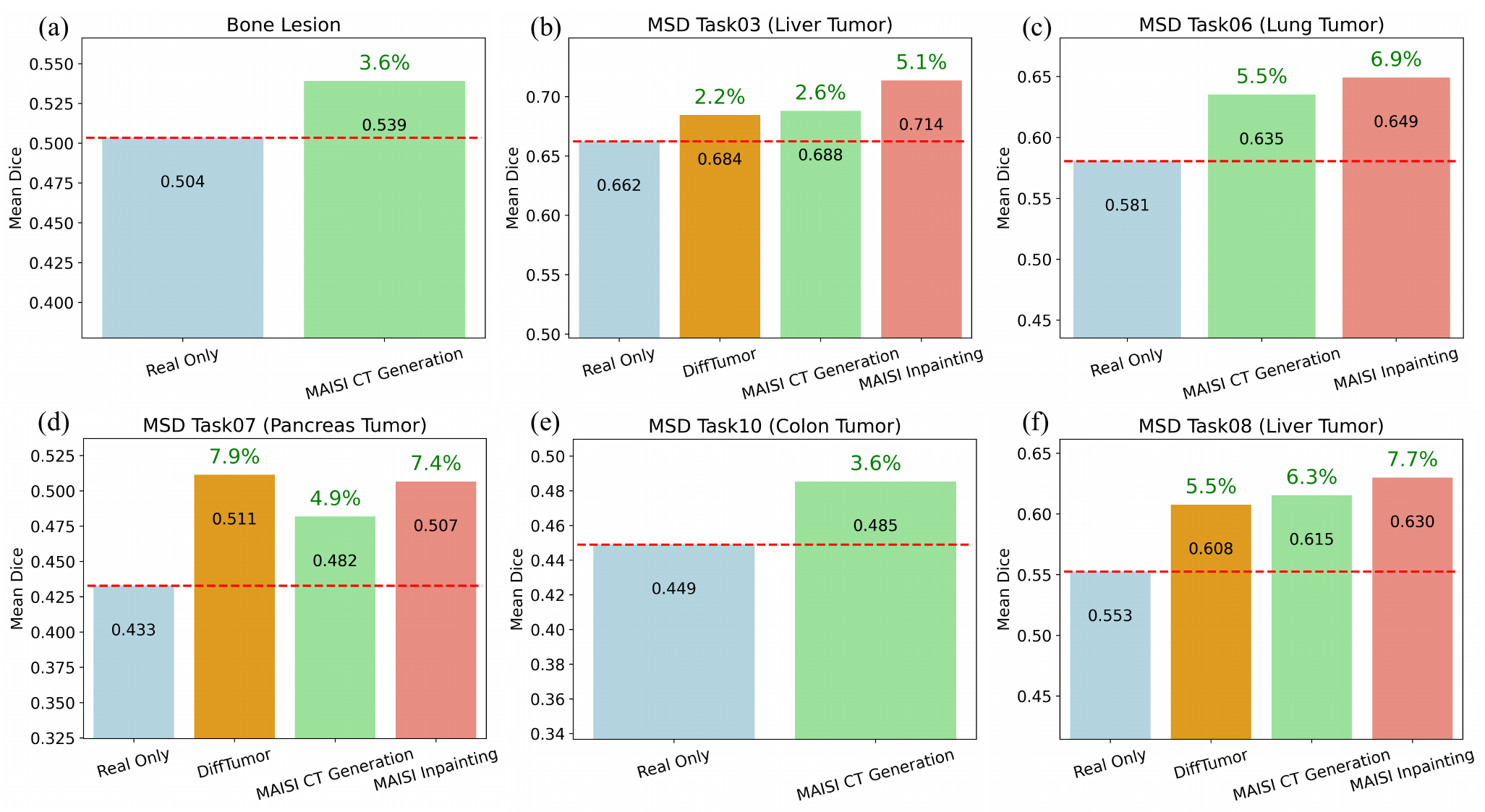}
\caption{The 5-fold averaged DSC of data augmentation experiments using synthetic data across 5 tumor types. The percentage of relative improvement compared to \textbf{Real Only} experiments is shown in green above each bar plot. All reported improvements are significant under the Wilcoxon signed rank test.}
\label{fig:data_aug_exps}
\end{figure*}

 One of the critical applications of generative models in medical imaging is data augmentation for training deep learning models. To assess the effectiveness of synthetic images in improving model performance, especially for rare medical conditions, we integrate synthetic data generated by MAISI into a standard training pipeline and evaluate it across five tumor types. Specifically, we employ the Auto3DSeg\footnote{https://monai.io/apps/auto3dseg} pipeline—an auto-configuration solution for training medical image segmentation models—to train models on the MSD Task03~\cite{antonelli2022medical} (liver tumor), Task06~\cite{antonelli2022medical} (lung tumor), Task07~\cite{antonelli2022medical} (pancreas tumor), Task10~\cite{antonelli2022medical} (colon tumor), and an in-house bone lesion dataset. We conduct experiments by training segmentation models either using only real data (referred to \textbf{Real Only} in Fig.~\ref{fig:data_aug_exps}) or by incorporating synthetic data from different models, thereby demonstrating the impact of synthetic data on data augmentation. To ensure robustness, we performed 5-fold cross-validation and reported the average Dice Similarity Coefficient (DSC) on the testing set across the five folds. 
 
 As discussed in Sec.\ref{sec: controlnet}, the integration of ControlNet~\cite{zhang2023adding} introduces a flexible mechanism in MAISI, enabling the incorporation of task-specific conditions. To illustrate its versatility, we trained ControlNet~\cite{zhang2023adding} for two distinct tasks aimed at generating synthetic data for augmentation purposes. The first task (denoted as \textbf{MAISI CT Generation} in Fig.~\ref{fig:data_aug_exps}) is conditional generation from segmentation masks of 127 anatomical structures, including the five tumor types mentioned earlier. This approach allowed us to generate synthetic data by augmenting real patient tumor masks corresponding to each tumor type. The second task (denoted as \textbf{MAISI Inpainting} in Fig.~\ref{fig:data_aug_exps}) involves training a tumor inpainting model designed to simultaneously support liver, pancreas, and lung tumors, following the setting in\cite{chen2024towards}. The tumor inpainting model requires a function to simulate tumor masks for adding synthetic tumors into healthy patient data. However, simulating tumors with irregular shapes, such as bone lesions and colon tumors, poses significant challenges. For a comparative analysis, we benchmark against the state-of-the-art tumor synthesis method,
DiffTumor~\cite{chen2024towards}, using their released model\footnote{https://github.com/MrGiovanni/DiffTumor} which supports liver and pancreas tumors among five tumor types in our experiments. 

Results shown in Fig.~\ref{fig:data_aug_exps}(a)$\sim$(e) indicate prominent improvements in DSC scores across all tumor types when incorporating synthetic data from our two augmentation tasks. Specifically, the MAISI CT Generation results in an average DSC improvement of 4\% across the five tumor types. The MAISI Inpainting demonstrated a more substantial average improvement of 6.5\% in DSC for liver, lung, and pancreas tumors, performing comparably or better than the DiffTumor~\cite{chen2024towards}, which trains dedicated synthesis models for each tumor type. Additionally, we conduct an out-of-distribution evaluation by testing tumor segmentation models trained on MSD Task03~\cite{antonelli2022medical} on 303 liver tumor samples from MSD Task08~\cite{antonelli2022medical}. As shown in Fig.~\ref{fig:data_aug_exps}(f), models incorporating synthetic data consistently show greater relative performance improvements compared to those evaluated within their original training dataset in Fig.~\ref{fig:data_aug_exps}(b). These findings underscore the effectiveness of synthetic data as a powerful augmentation strategy to bolster the generalizability of segmentation models. More ablation studies and visualization of synthetic data can be found in Supplementary Sec.~\ref{supp: exp}.

\section{Discussion and Limitations}
While the proposed MAISI demonstrates great potentials in generating high-quality CT images, it is essential to recognize its limitations and potential societal impacts. While MAISI shows robust performance across various datasets, its ability to accurately represent demographic variations (such as age, ethnicity, and gender differences) in generated anatomy has not been extensively validated. Future studies can focus on ensuring that synthetic data adequately captures this diversity to avoid bias in downstream applications. The capabilities of generating high-resolution images of MAISI, while innovative, still demand substantial computation resources. This could limit accessibility for researchers and institutions with less computational power, potentially widening the gap between high-resource and low-resource entities. Future efforts can focus on improving the accessibility of MAISI, particularly in resource-constrained environments.

\section{Conclusion}
In this paper, we propose MAISI, a novel framework for generating high-resolution 3D CT volumes using a combination of foundation models and ControlNet~\cite{zhang2023adding}. MAISI aims to provide an adaptable and versatile solution for generating anatomically accurate images. Our experiments demonstrate that MAISI can produce realistic CT images with flexible volume dimensions and voxel spacing, offering promising potential to augment medical datasets and improve the performance of downstream tasks.

{\small
\bibliographystyle{ieee_fullname}
\bibliography{egbib}
}
\clearpage
\onecolumn    
\appendix
\beginsupplement

This supplementary material is organized as follows: Sec.~\ref{supp: data} provides more details about the datasets utilized in model training. More implantation details about three networks and downstream tumor segmentation tasks are provided in Sec.~\ref{supp: training}. Sec.~\ref{supp: exp} contains additional visualizations of synthetic data and ablation studies.

\section{Dataset Details}\label{supp: data}
\subsection{MAISI VAE}
For the foundational 3D VAE in MAISI, we include a diverse dataset comprising 37,243 CT volumes for training and 1,963 CT volumes for validation, covering the chest, abdomen, and head and neck regions. Additionally, we include 17,887 MRI volumes for training and 940 MRI volumes for validation, spanning the brain, skull-stripped brain, chest, and below-abdomen regions. The training data were sourced from various repositories, including TCIA COVID-19 Chest CT, TCIA Colon Abdomen CT, MSD03 Liver Abdomen CT, LIDC Chest CT, TCIA Stony Brook COVID Chest CT, NLST Chest CT, TCIA Upenn GBM Brain MR, AOMIC Brain MR, QTIM Brain MR, TCIA Acrin Chest MR, and TCIA Prostate MR. This extensive and varied dataset not only ensures that our model is exposed to a broad range of anatomical regions but also supports its application to both MRI and CT images.

The details of MAISI VAE training data are shown in Table~\ref{tab:vae_data}.
\begin{table}[ht]
\centering
\begin{tabular}{|l|c|c|}
\hline
\textbf{Dataset Name} & \textbf{Number of Training Data} & \textbf{Number of Validation Data} \\
\hline
Covid 19 Chest CT & 722 & 49 \\
TCIA Colon Abdomen CT & 1522 & 77 \\
MSD03 Liver Abdomen CT & 104 & 0 \\
LIDC chest CT & 450 & 24 \\
TCIA Stony Brook Covid Chest CT & 2644 & 139 \\
NLST Chest CT & 31801 & 1674 \\
TCIA Upenn GBM Brain MR (skull-stripped) & 2550 & 134 \\
Aomic Brain MR & 2630 & 138 \\
QTIM Brain MR & 1275 & 67 \\
Acrin Chest MR & 6599 & 347 \\
TCIA Prostate MR Below-Abdomen MR & 928 & 49 \\
Aomic Brain MR, skull-stripped & 2630 & 138 \\
QTIM Brain MR, skull-stripped & 1275 & 67 \\
\hline
\textbf{Total CT} & 37243 & 1963 \\
\textbf{Total MRI} & 17887 & 940 \\
\hline
\end{tabular}
\caption{MAISI VAE Dataset Information}
\label{tab:vae_data}
\end{table}

\subsection{MAISI Diffusion}

\begin{figure}[ht] 
\centering
\includegraphics[width=\columnwidth]{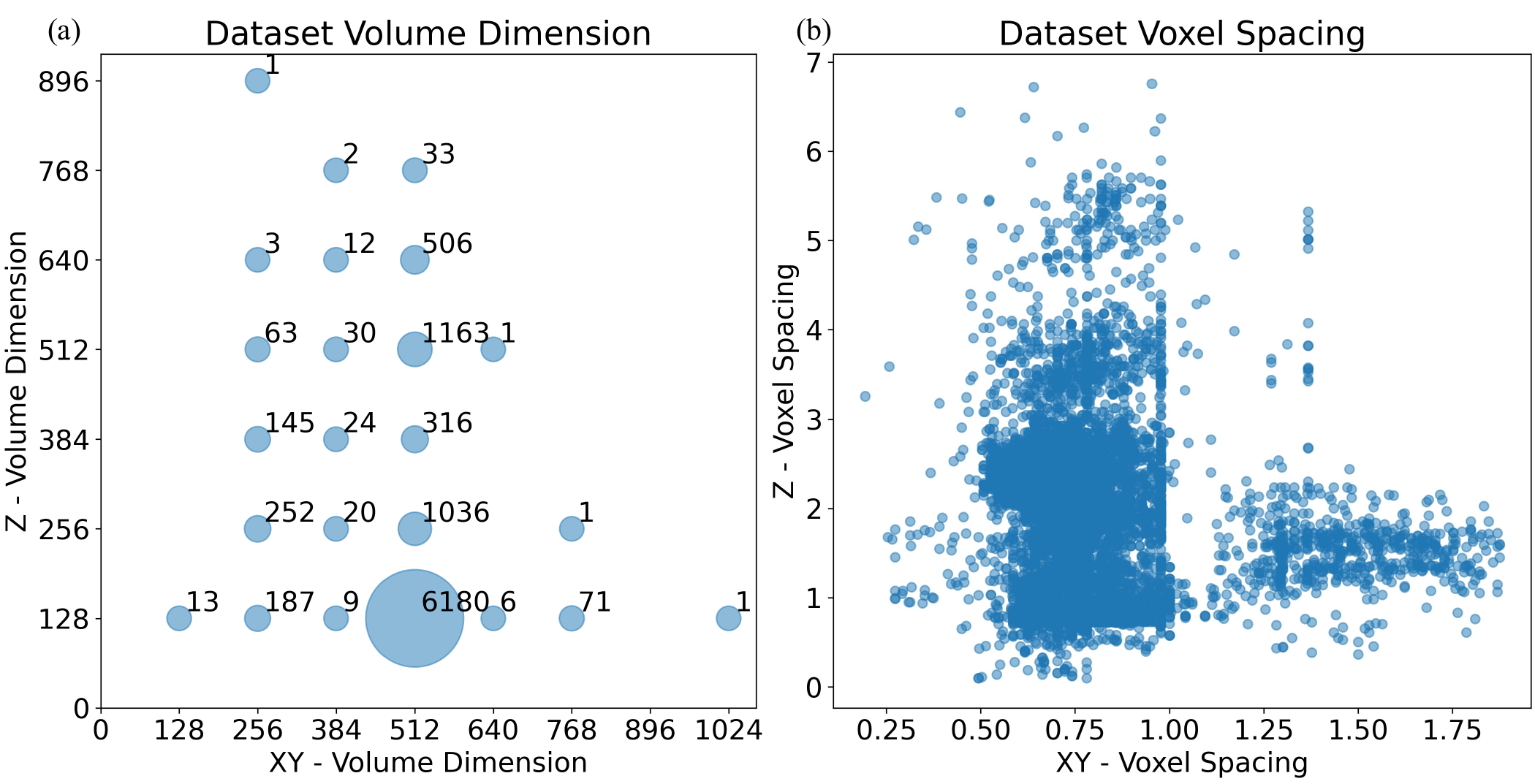}
\caption{The characteristics of the datasets utilized for the MAISI Diffusion Model are detailed through two subplots. Subplot (a) illustrates the volume dimensions of the datasets, providing insight into the variability and range of sizes used in the training data. Subplot (b) presents the voxel spacing in millimeters for each data point, emphasizing the spatial configuration within the CT scans. Notably, in CT imaging, the X and Y directions typically share identical dimensions and spacing, so they are represented on a single axis in both subplots. }
\label{fig:dataset_info}
\end{figure}

The datasets for developing the Diffusion model used in MAISI comprise 10,277 CT volumes from 24 distinct datasets, encompassing various body regions and disease patterns. Table~\ref{tab:dm_data} provides a summary of the number of volumes for each dataset. For compatibility with the shape requirement of the U-shape network, we resample the dimensions of volumes to multiples of 128. Fig.~\ref{fig:dataset_info} visualizes the characteristics and spatial complexity of the data involved in training the diffusion model. 
\begin{table}[ht]
\centering
\begin{tabular}{|l|c|}
\hline
\textbf{Dataset name}    & \textbf{Number of volumes} \\ \hline
AbdomenCT-1K             & 789                        \\
AeroPath                 & 15                         \\
AMOS22                   & 240                        \\
autoPET23 (testing only)                & 200                        \\
Bone-Lesion              & 223                        \\
BTCV                     & 48                         \\
COVID-19                 & 524                        \\
CRLM-CT                  & 158                        \\
CT-ORG                   & 94                         \\
CTPelvic1K-CLINIC        & 94                         \\
LIDC                     & 422                        \\
MSD Task03               & 88                         \\
MSD Task06               & 50                         \\
MSD Task07               & 224                        \\
MSD Task08               & 235                        \\
MSD Task09               & 33                         \\
MSD Task10               & 87                         \\
Multi-organ-Abdominal-CT & 65                         \\
NLST                     & 3109                       \\
Pancreas-CT              & 51                         \\
StonyBrook-CT            & 1258                       \\
TCIA\_Colon              & 1437                       \\
TotalSegmentatorV2       & 654                        \\
VerSe                    & 179                        \\ \hline 
\textbf{Total}           & 10277                        \\ \hline 
\end{tabular}
\caption{MAISI DM Dataset Information}
\label{tab:dm_data}
\end{table}

\subsection{MAISI ControlNet}

The ControlNet training dataset for \textbf{MAISI CT Generation} discussed in Sec.~\ref{sec: da} contains 6,330 CT volumes (5,058 and 1,272 volumes are used for training and validation, respectively) across 20 datasets and covers different body regions and diseases.
Table~\ref{tab:controlnet_data} summarizes the number of volumes for each dataset.
\begin{table}[ht]
\centering
\begin{tabular}{|l|c|}
\hline
\textbf{Dataset name}    & \textbf{Number of volumes} \\ \hline
AbdomenCT-1K             & 789                        \\
AeroPath                 & 15                         \\
AMOS22                   & 240                        \\
Bone-Lesion              & 237                        \\
BTCV                     & 48                         \\
CT-ORG                   & 94                         \\
CTPelvic1K-CLINIC        & 94                         \\
LIDC                     & 422                        \\
MSD Task03               & 105                        \\
MSD Task06               & 50                         \\
MSD Task07               & 225                        \\
MSD Task08               & 235                        \\
MSD Task09               & 33                         \\
MSD Task10               & 101                        \\
Multi-organ-Abdominal-CT & 64                         \\
Pancreas-CT              & 51                         \\
StonyBrook-CT            & 1258                       \\
TCIA\_Colon              & 1436                       \\
TotalSegmentatorV2       & 654                        \\
VerSe                    & 179                        \\ \hline
Total                    & 6330                       \\ \hline
\end{tabular}
\caption{MAISI ControlNet Dataset Information}
\label{tab:controlnet_data}
\end{table}

\clearpage
\section{Additional Implementation Details}\label{supp: training}
\noindent\textbf{MAISI VAE.}
To establish the VAE as a foundational model, we employ an extensive range of data augmentation techniques. For CT images, intensities are clipped to a Hounsfield Unit (HU) range of -1000 to 1000 and normalized to a range of [0,1]. For MR images, intensities were normalized such that the 0th to 99.5th percentile values were scaled to the range [0,1]. For MR images, we applied intensity augmentations including random bias field, random Gibbs noise, random contrast adjustment, and random histogram shifts. Both CT and MR images underwent spatial augmentations, such as random flipping, random rotation, random intensity scaling, random intensity shifting, and random upsampling or downsampling.

The MAISI VAE model is trained with 8 32G V100 GPU. It is initially trained for 100 epochs using small, randomly cropped patches of size [64,64,64]. This approach is adopted to improve the model's ability to generalize to images with partial volume effects. After this initial phase, training is continued for an additional 200 epochs using larger patches of size [128,128,128], which allows the model to capture more contextual information and improve overall accuracy.

The MAISI VAE is used to compress the latent features that will be employed in latent diffusion models, where having a well-structured and meaningful latent space is crucial for effective diffusion dynamics. Therefore, during MAISI VAE training, we adjust the weight of the KL loss to ensure the standard deviation remains between 0.9 to 1.1. This calibration balances the model’s focus between accurate data reconstruction and adherence to the prior distribution. As the MAISI VAE is intended to serve as a foundational model, maintaining this balance also helps to prevent over-fitting \cite{higgins2017beta}.

\noindent\textbf{MAISI Diffusion.}
Data preprocessing for diffusion model training involves applying a series of precise transformations to the image data, including loading the images, ensuring the correct channel structure, adjusting the orientation according to the "RAS" axcode, and scaling intensity values from $-1000$ to $1000$ to normalize the data between 0 and 1.
The process further refines the images by adjusting dimensions to the nearest multiple of $128$, recording the new spatial details, using trilinear interpolation.
Then each image is passed through a pre-trained autoencoder, generating a compressed latent representation that is saved for subsequent model training.
The diffusion model requires additional input attributes, including output dimensions, output spacing, and top/bottom body region indicators. These dimensions and spacing are extracted from the header information of the training images. The top and bottom body regions can be identified either through manual inspection or by using segmentation tools such as TotalSegmentator~\cite{wasserthal2023totalsegmentator} and VISTA3D~\cite{he2024vista3d}. These regions are encoded as $4$-dimensional one-hot vectors: the head and neck region is represented by $[1,0,0,0]$, the chest by $[0,1,0,0]$, the abdomen by $[0,0,1,0]$, and the lower body (below the abdomen) by $[0,0,0,1]$. These additional input attributes are stored in a separate configuration file. In this example, it is assumed that the images encompass the chest and abdomen regions.

Next, the diffusion model training process begins with an initial learning rate of $1e^{-4}$, a batch size of $1$, and spans $200$ epochs. To ensure the data is optimally prepared for training, various transformations are applied to the image inputs.
The U-Net architecture is employed for noise prediction, with distributed computing utilized to enhance efficiency when multiple GPUs are available.
The Adam optimizer is responsible for adjusting the model's parameters, while a polynomial learning rate scheduler controls the update rate over training steps.
Noise is systematically introduced to the input data by the noise scheduler, and the model iteratively refines its predictions using an L1 loss function to minimize this noise.
Mixed precision training and gradient scaling are implemented to optimize memory usage and computational performance.

\noindent\textbf{MAISI ControlNet.} We train a versatile ControlNet Model (MAISI CT Generation task in Sec.~\ref{sec: da}) to support all five types of tumors using the datasets summarized in Table~\ref{tab:controlnet_data}. The data preprocessing protocol is the same in the training of the MAISI Diffusion Model. The Adam optimizer is employed for training purposes, with hyperparameters $\beta_1=0.9$ and $\beta_2=0.999$. The learning rate is set at 0.0001, with the polynomial learning rate decay. The batch size is set to 1 per GPU. Training is performed on a server with 8 A100 GPUs with about 10k optimization steps. For the MAISI Inpainting task, we employ the same hyperparameters for training but only use datasets with supported tumor types, including MSD Task03~\cite{antonelli2022medical} (liver tumor), Task06~\cite{antonelli2022medical} (lung tumor), Task07~\cite{antonelli2022medical} (pancreas tumor).

\begin{figure}[ht] 
\centering
\includegraphics[width=\columnwidth]{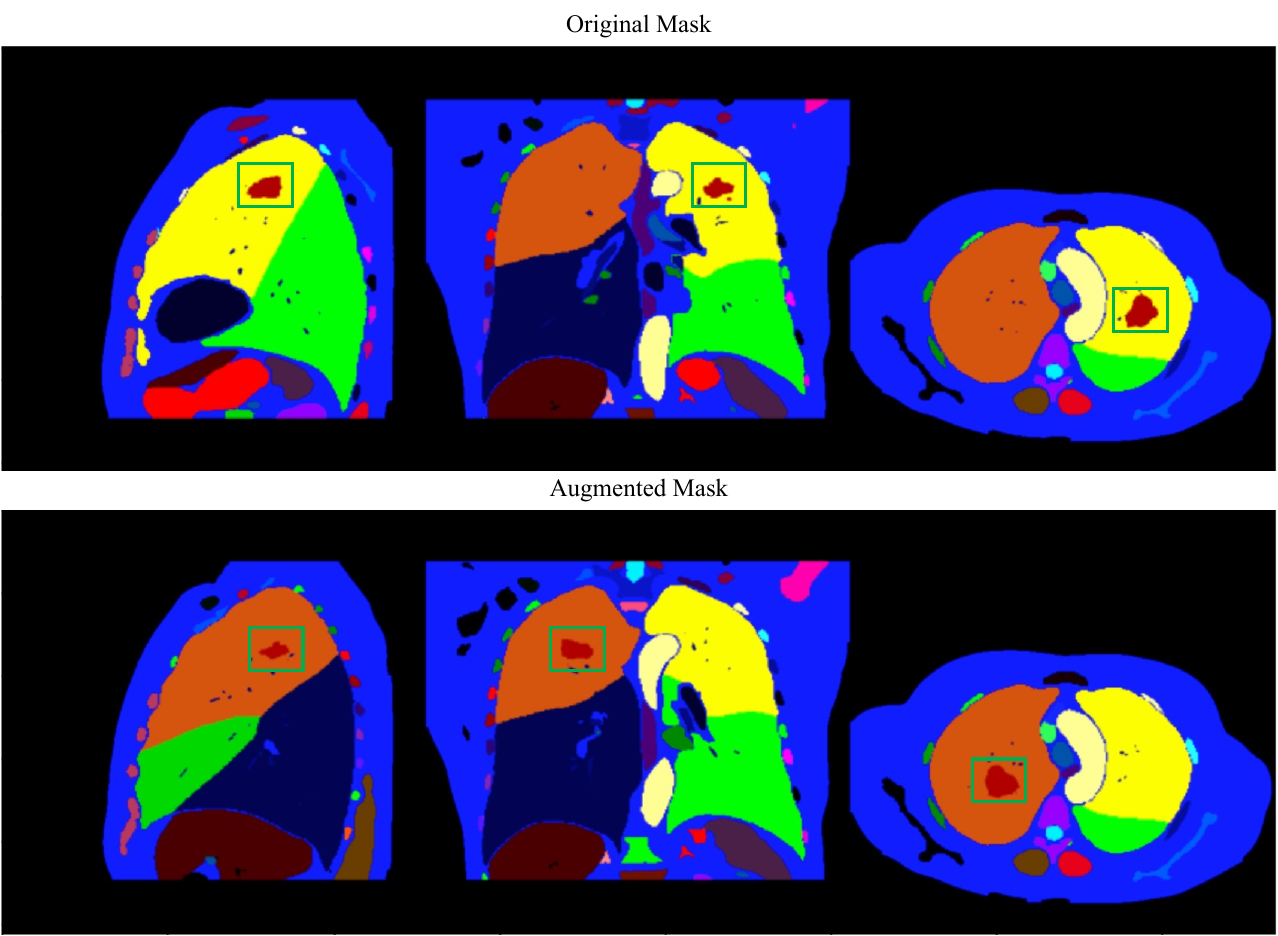}
\caption{The example lung tumor mask and corresponding augmented mask. The green boxes highlight the tumor regions in different views.}
\label{fig:mask_aug}
\end{figure}

\noindent\textbf{Downstream tumor segmentation.} The implementation of all tumor segmentation models is based on the Auto3DSeg\footnote{https://monai.io/apps/auto3dseg} pipeline. Auto3DSeg is an auto-configuration pipeline designed for 3D medical image segmentation, utilizing MONAI~\cite{cardoso2022monai}. The pipeline begins with data analysis to extract global information from the dataset, followed by algorithm generation based on data statistics and predefined templates. It then proceeds to model training to obtain optimal checkpoints. All used tumor dataset is split into 80\% for training and 20\% for testing. The training set is further divided into five folds for 5-fold cross-validation. We report the segmentation performance on the holdout testing set. For the MAISI CT Generation task, we generate synthetic data from augmented real masks containing tumors. Fig.~\ref{fig:mask_aug} shows an example of mask augmentation for a case with the lung tumor. For the MAISI Inpainting task, we follow the same setting in DiffTumor~\cite{chen2024towards} and use the provided healthy cases in the open-source repository\footnote{https://github.com/MrGiovanni/DiffTumor} to generate synthetic data with tumors. For both tasks, the amount of synthesized data is equivalent to the original dataset size for each tumor type. We explore the impact of using different amounts of synthetic data for data augmentation in Supplementary Sec.~\ref{supp: exp}.

\clearpage

\section{Supplementary Experiment Results}\label{supp: exp}

\begin{figure}[ht] 
\centering
\includegraphics[width=\columnwidth]{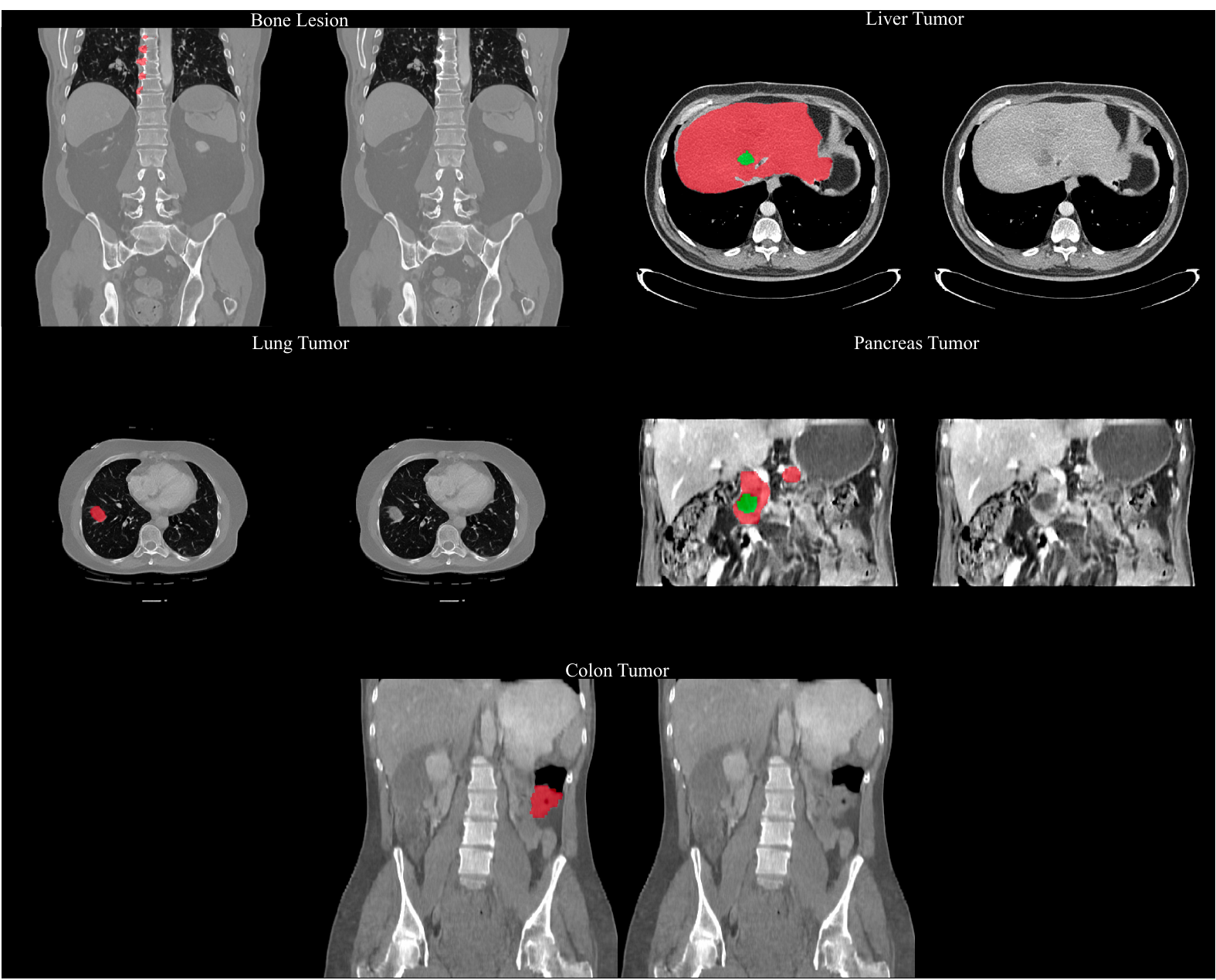}
\caption{The example of generated images from MAISI CT Generation task.}
\label{fig:example_maisi_ct_gen}
\end{figure}

\begin{figure}[ht] 
\centering
\includegraphics[width=0.65\columnwidth]{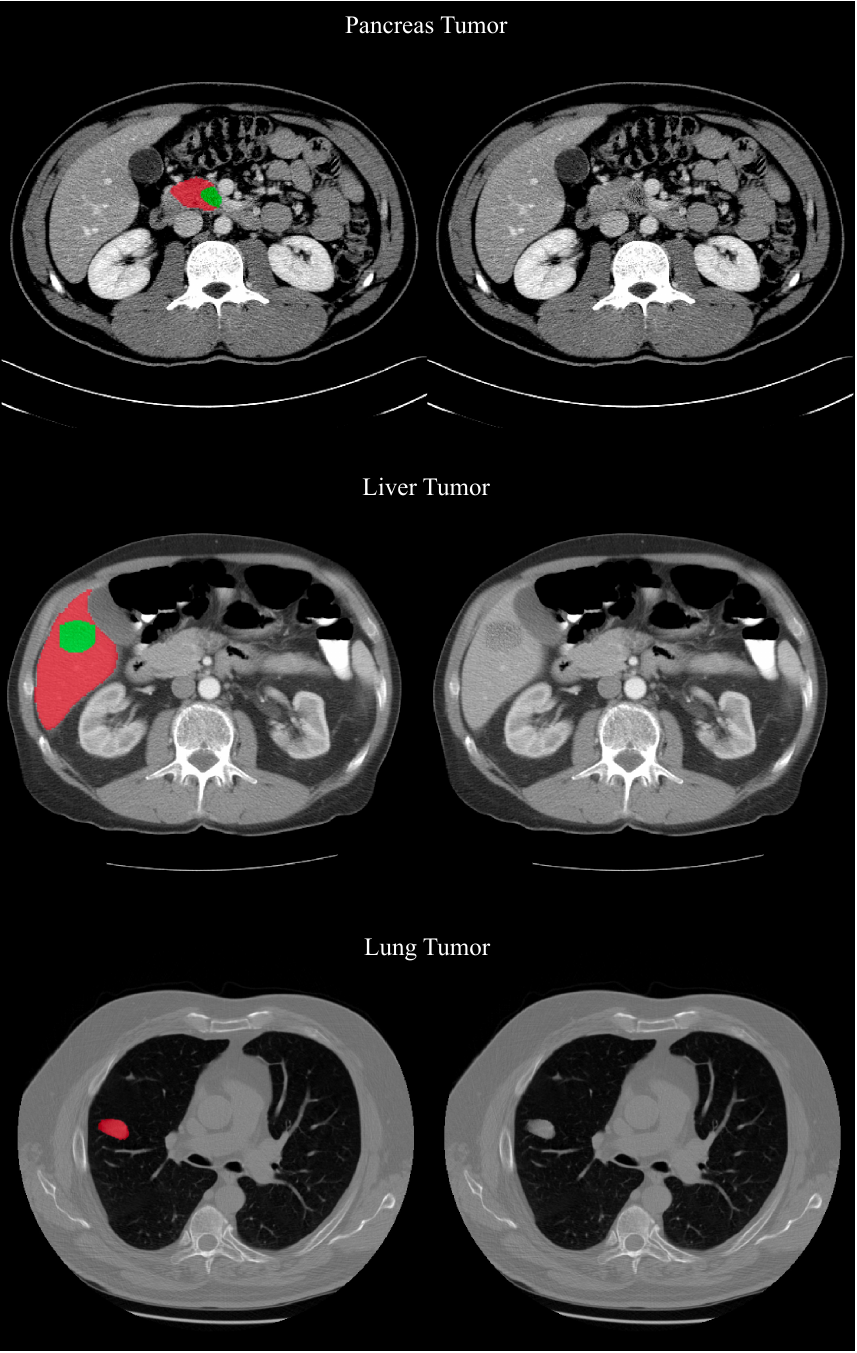}
\caption{The example of generated images from MAISI Inpainting task.}
\label{fig:example_maisi_inpainting}
\end{figure}

\begin{table}[ht!]
\small
\resizebox{0.99\columnwidth}{!}{
\begin{tabular}{c|c|cccccc|c}
\hline
MSD Task06          & Real v.s. Synthetic & fold 0 & fold 1 & fold 2 & fold 3 & fold 4 & Avg.  & Improvement \\ \hline\hline
Real Only           & 1:0                 & 0.494  & 0.601  & 0.535  & 0.674  & 0.599  & 0.581 & -           \\
MAISI CT Generation & 1:1                 & 0.585  & 0.649  & 0.631  & 0.647  & 0.664  & 0.635 & 5.5\%       \\
MAISI CT Generation & 1:0.5               & 0.640  & 0.593  & 0.606  & 0.639  & 0.644  & 0.624 & 4.4\%       \\
MAISI CT Generation & 1:1.5               & 0.641  & 0.658  & 0.586  & 0.645  & 0.666  & 0.639 & 5.8\%       \\ \hline
MSD Task07          & Real v.s. Synthetic & fold 0 & fold 1 & fold 2 & fold 3 & fold 4 & Avg.  & Improvement \\ \hline\hline
Real Only           & 1:0                 & 0.423  & 0.463  & 0.414  & 0.42   & 0.444  & 0.433 & -           \\
MAISI CT Generation & 1:1                 & 0.504  & 0.448  & 0.467  & 0.482  & 0.508  & 0.482 & 4.9\%       \\
MAISI CT Generation & 1:0.5               & 0.465  & 0.463  & 0.423  & 0.447  & 0.478  & 0.455 & 2.2\%       \\
MAISI CT Generation & 1:1.5               & 0.466  & 0.481  & 0.465  & 0.480  & 0.467  & 0.471 & 3.9\%       \\ \hline
\end{tabular}
}
\caption{The ablation study examines the effect of varying amounts of synthetic data in data augmentation experiments. The 'Improvement' column reports the percentage of relative improvement compared to experiments using only real data. We conduct this ablation study on the smallest dataset (MSD Task06) and the largest dataset (MSD Task07) across five tumor types. Our empirical results suggest that using a synthetic dataset equivalent in size to the original dataset is an effective choice for data augmentation.
}
\label{tab:ablation_ratio}
\end{table}
\begin{table}[ht!]
\small
\resizebox{0.99\columnwidth}{!}{
\begin{tabular}{c|cccccccccccc}
\hline
\multicolumn{1}{l|}{} & Liver & Spleen & Left Kidney & Right Kidney & Stomach & Gallbladder & Esophagus & Pancreas & Duodenum & Colon & Small Bowel & Bladder \\ \hline
Real Data             & 0.95  & 0.94   & 0.93        & 0.93         & 0.90    & 0.75        & 0.76      & 0.80     & 0.69     & 0.76  & 0.80        & 0.91    \\
Synthetic Data        & 0.93  & 0.93   & 0.95        & 0.95         & 0.88    & 0.47        & 0.73      & 0.70     & 0.54     & 0.73  & 0.74        & 0.86    \\ \hline
\end{tabular}
}
\caption{Segmentation performance on synthetic data. Synthetic data is generated using the MAISI CT Generation task and evaluated with the VISTA 3D~\cite{he2024vista3d} segmentation model. DSC are presented for both synthetic and real data on the unseen WORD~\cite{luo2022word} dataset. The results demonstrate that the segmentation model achieves comparable performance on major organs (\eg, liver, spleen, kidney) for both synthetic and real data. However, smaller organs (\eg, gallbladder, duodenum, pancreas) show a more pronounced performance gap between synthetic and real data. Addressing this gap presents a promising direction for future research.}
\label{tab:ablation_seg}
\end{table}
\end{document}